\documentclass[12pt,a4paper]{article}

\usepackage{amssymb}
\usepackage{amsmath}
\usepackage{mathrsfs}

\usepackage{cite}

\allowdisplaybreaks[1]

\DeclareMathOperator*{\tr}{tr}

\begin{document}

\thispagestyle{empty}

\begin{flushright}
 KEK-TH-1257
\end{flushright}

\vspace{10mm}
\begin{center}
{\Large \bf Scaling limit of ${\cal N}=6 $
superconformal Chern-Simons  theories and 
Lorentzian Bagger-Lambert theories}

\vspace{10mm}
{\large
Yoshinori Honma\footnote{E-mail address: 
yhonma@post.kek.jp},
Satoshi Iso\footnote{E-mail address: satoshi.iso@kek.jp},
Yoske Sumitomo\footnote{E-mail address: sumitomo@post.kek.jp}
and Sen Zhang\footnote{E-mail address: zhangsen@post.kek.jp}
}

\vspace{10mm}
 {\it Institute of Particle and Nuclear Studies, \\ High Energy Accelerator Research Organization(KEK) \\
 and \\ 
 Department of Particles and Nuclear Physics, \\
 The Graduate University for Advanced Studies (SOKENDAI), 
 \vspace{3mm} \\
Oho 1-1, Tsukuba, Ibaraki 305-0801, Japan} 

\end{center}

\vspace{20mm}
\ \ 

\begin{abstract}
We show that the
${\cal N}=8$ superconformal Bagger-Lambert theory
based on the Lorentzian 3-algebra can be
derived by taking a certain scaling limit
of the  recently proposed
${\cal N}=6$ superconformal
$U(N) \times U(N)$
Chern-Simons-matter theories at level
$(k, -k)$.
The scaling limit (and In\"on\"u-Wigner contraction) is
to scale the trace part of the bifundamental 
fields as $X_0 \rightarrow \lambda^{-1} X_0$
and an axial combination of the two gauge fields as 
$B_{\mu} \rightarrow \lambda B_\mu$.
Simultaneously we scale the level 
as $k \rightarrow \lambda^{-1} k$ and then
take $\lambda \rightarrow 0$ limit.
Interestingly the same constraint equation
$\partial^2 X_0=0$ is
 derived by imposing finiteness of the action.
In this scaling limit, M2-branes are located
far from the origin of ${\bf C^4 /Z_k}$
compared to their fluctuations and ${\bf Z_k}$
identification becomes a circle identification.
Hence the scaled theory describes  ${\cal N}=8$
supersymmetric theory of 2-branes  
with dynamical  coupling.
The coupling constant  is promoted to
a space-time dependent $SO(8)$ vector $X_0^I$ and 
we show that  the scaled theory  has 
 a {\it generalized conformal symmetry}
as well as manifest $SO(8)$ with the transformation of the background fields $X_0^I$. 

\end{abstract}

\newpage
\setcounter{page}{1}
\setcounter{footnote}{0}

\baselineskip 6mm
\section{Introduction}
\setcounter{equation}{0}
Recently there has been a lot of activity
in constructing superconformal Chern-Simons-matter
gauge theories for multiple M2-branes.
Bagger and Lambert, and Gustavsson
discovered the ${\cal N}=8$ superconformal
$(2+1)$-dimensional field theories with $SO(8)$
global symmetries by exploiting the 3-algebra
structures \cite{Bagger:2007jr,Gustavsson:2007vu}. 

The 3-algebra, or the quantization of the Nambu bracket,
is a natural generalization of the ordinary Lie algebras
and  has been widely believed that it will play an 
important role in the quantization of membranes
\cite{Nambu,Awata,Kawamura,O'Farrill,Ho:2008bn,Papa}.
Furthermore the 3-algebraic structure is also 
expected to play an important role to construct
M5 branes from M2-branes through the Basu-Harvey 
equation \cite{Basu:2004ed}.

The quantization of the Nambu
bracket is  hard, if we impose the so-called
Fundamental Identity (FI) on the 3-algebra
and the antisymmetry of the structure constants,
and there are only a few examples. 
The simplest one is  the ${\cal A}_4$ algebra with
four generators. The Bagger-Lambert theory based
on the ${\cal A}_4$ algebra becomes an 
${\cal N}=8$ superconformal $SU(2) \times SU(2)$ 
Chern-Simons gauge theory \cite{VanRaamsdonk:2008ft}.
By giving a vacuum expectation value (VEV) to the matter field, the theory
is shown to describe low energy effective theory
of D2-branes\cite{Mukhi:2008ux}. 

Another remarkable example is 
the 3-algebra  with a Lorentzian metric \cite{Gomis:2008uv,Benvenuti:2008bt,Ho:2008ei}.
It satisfies the FI and 
 ${\cal N}=8$ superconformal field
theories based on this Lorentzian 3-algebra are constructed.
Because of the Lorentzian signature,
the model contains ghost fields and 
they may break the unitarity of the theory.
The fields with a negative signature, however,
are Lagrange multipliers and can be
 integrated out to give constraints on 
the ``conjugate'' fields. 
After the integration, the theory looks
well-defined. (There are also attempts
to kill the ghost fields by gauging shift symmetry  \cite{Bandres:2008kj,Gomis:2008be}.)
If we take a special solution to the 
constraint equation, the Lorentzian  
Bagger-Lambert theories are 
reduced to the action of the $N$ D2-branes
in flat space.  
There are various solutions to the constraint
equations and each of them give different looking
Janus field theories \cite{Bak:2003jk,D'Hoker:2006uv,Kim:2008dj,Gaiotto:2008sd,D'Hoker:2008wc} of D2-branes whose coupling
is varying with the space-time \cite{Honma:2008un}.

There are various generalizations of the BL theories,
including massive deformations \cite{Gomis:2008cv,Hosomichi:2008qk,Song:2008bi}
and models based on 
other 3-algebras\cite{Lin:2008qp}. These theories are also
obtained from gauged supergravities in three dimensions
by using the embedding tensor method \cite{embedding}.
For other related works, see
\cite{others,others3,others4,others5,others6,Gauntlett:2008uf,Bandres:2008vf,Papadopoulos:2008gh,others8,Fuji:2008yj,Ho:2008ve,Krishnan:2008zm,Jeon:2008bx,Li:2008ez,Hosomichi:2008jd,Banerjee:2008pd,FigueroaO'Farrill:2008zm,Gustavsson:2008bf,Ahn:2008ya,Passerini:2008qt,Ezhuthachan:2008ch,Cecotti:2008qs,Mauri:2008ai,deMedeiros:2008bf,Blau:2008bm}.

Another very interesting proposal  for multiple
M2-branes actions was recently given by Aharony, Bergman, Jafferis and
Maldacena (ABJM) \cite{Aharony:2008ug}. They generalized the
superconformal Chern-Simons matter theories 
\cite{Schwarz:2004yj,Gaiotto:2007qi} to the ${\cal N}=6$
superconformal $U(N) \times U(N)$ theories.
The level of the Chern-Simons gauge theories
is $(k,-k)$ and the theory is conjectured to
 describe the low energy limit of $N$ M2-branes
 probing a ${\bf C^4/Z_k}$. Hence at large $N$, it is 
dual to the M-theory on 
$AdS_4 \times S^7/{\bf Z_k}$.
In the case of $SU(2) \times SU(2)$ gauge group,
the theory is the same as the Bagger-Lambert
theory based on the ${\cal A}_4$ 3-algebra\cite{VanRaamsdonk:2008ft}.
In this formulation by Aharony et.al.
 the 3-algebra structure does not seem to play 
 any role and, for the general gauge groups, 
the relation to the BL theory is not clear.
However, since
the theory with a gauge group $U(N) \times U(N)$
is conjectured to   describe $N$ M2-branes
probing ${\bf C^4/Z_k}$ and
giving a VEV to the bifundamental field 
reduces the theory to a system of $N$ D2-branes,   
it must be related, by taking a certain scaling limit, to 
the Lorentzian Bagger-Lambert theory with 
gauge group $U(N)$  which can be 
also reduced to the system of $N$ D2-branes
with a dynamical coupling.

There are various studies of
 the ABJM theory including an orbifolding
\cite{Benna:2008zy},  calculation of  the index \cite{Bhattacharya:2008bj},
and PP wave limit \cite{PPCS}.

In this letter, we show that the Lorentzian 
Bagger-Lambert theory can be obtained by
taking an appropriate scaling limit
of the ABJM theory.
We first scale the trace of the bifundamental
fields $X_0$ (bosons and fermions)
and an axial combination of the
gauge fields $B_\mu=(A^{(L)}_\mu - A_\mu^{(R)})/2$ 
as
\begin{equation}
X_0 \rightarrow \lambda^{-1} X_0, \ \ 
B_\mu \rightarrow \lambda B_\mu.
\end{equation}
The other fields are kept fixed.
Simultaneously we scale $k \rightarrow \lambda^{-1} k$
and then  take $\lambda \rightarrow 0$ limit (In\"on\"u-Wigner
contraction \cite{Inonu:1953sp})\footnote{This scaling limit was also
discussed to get the nonsemisimple algebra in \cite{Tseytlin:1995yw}.}.
Many terms vanish in this limit, and the
remaining terms give the action of
the Lorentzian Bagger-Lambert theory
whose Lagrange multiplier fields are integrated out.
Interestingly divergent terms in 
$\lambda \rightarrow 0$ limit give constraints 
on the trace component of the bifundamental fields $X_0$.
The constraints precisely agree with the 
constraints given by the Lagrange multiplier
fields in the Lorentzian Bagger-Lambert theory.

The M2-branes described by the ABJM theory are
expected to have  the conformal  and 
$SO(8)$ symmetries (for $k=1,2$).
The scaling limit we consider in our paper
corresponds to a limit of locating M2-branes
far from the origin of the ${\bf Z_k}$ orbifold
as well as taking $k \rightarrow \infty$.
Hence from the M2-brane point of view the scaled 
theory must have (a kind of) conformal symmetry
and (hidden) $SO(8)$ symmetry that the ordinary
D2-branes are not expected to have. 
Since the coupling constant of the 
scaled theory of D2-branes is promoted to a 
$SO(8)$ vector $X_0^I(x)$, we show that
the scaled theory has enhanced symmetries, i.e.
{\it generalized conformal symmetry} and 
$SO(8)$  invariance, if we allow $X_0^I (x)$ to transform appropriately under these transformations.
This generalized conformal
symmetry is essentially the same as that
proposed  by Jevicki, Kazama 
and Yoneya \cite{Jevicki} 10 years ago for 
general Dp-branes.

The paper is organized as follows.
In section 2, we compare the gauge structures
of the Lorentzian Bagger-Lambert theory
and the ABJM theory for M2-branes.
The gauge structure of the Lorentzian BL theory
is obtained by taking an In\"on\"u-Wigner contraction
of the ABJM theory.
In section 3, we look at the scaling limit of 
the ABJM theory and show that the scaled
action gives the Lorentzian Bagger-Lambert 
theory. We also see that the scaled theory
of D2-branes with a dynamical coupling has 
a generalized conformal symmetry.
We end in section 4 with conclusions and 
discussions.

\section{Gauge structures and In\"on\"u-Wigner contraction}
\setcounter{equation}{0}
We first look at the gauge structures of the
Lorentzian Bagger-Lambert theory \cite{Gomis:2008uv,Benvenuti:2008bt,Ho:2008ei}.
The Bagger-Lambert theory \cite{Bagger:2007jr,Gustavsson:2007vu} has a gauge symmetry
generated by $\tilde{T}^{ab} X = [T^a, T^b, X]$.
Because of the fundamental identity
\begin{align}
 [T^a,T^b,[T^c,T^d,T^e]] = [[T^a,T^b,T^c],T^d,T^e] +
  [T^c,[T^a,T^b,T^d],T^e] + [T^c,T^d,[T^a,T^b,T^e]],
\end{align}
the following commutation relation holds\footnote{
If we write the commutation relation
as $[\tilde{T}^{ab}, \tilde{T}^{cd}]=
f^{abc}_{ \ \ \ e} \tilde{T}^{ed} + 
f^{abd}_{\ \ \ e} \tilde{T}^{ce}$, it is 
not always associative. But when $\tilde{T}^{ab}$ acts
on a field $X$, associativity-violating terms 
(3-cocycles) vanish
and it becomes an ordinary associative Lie algebra.};
\begin{align}
 [\tilde{T}^{ab}, \tilde{T}^{cd}]X &= [T^a,T^b, [T^c,T^d,X]] - [T^c,T^d, [T^a,T^b,X]] 
\nonumber  \\ &= 
[[T^a,T^b,T^c],T^d,X]+ [T^c,[T^a,T^b,T^d], X]  
\nonumber \\
&= (f^{abc}_{ \ \ \ e} \tilde{T}^{ed} + f^{abd}_{\ \ \ e} \tilde{T}^{ce}) X.
\end{align}

The Lorentzian 3-algebra contains 
2 extra generators $T^{-1}$ and $T^0$
in addition to the generators of Lie algebra $T^i.$
(Here we use the convention of \cite{Ho:2008ei}.)
The 3-algebra for them is given by
\begin{align}
&[T^{-1}, T^a, T^b] = 0,  \label{Nb1}\\
&[T^0, T^i, T^j] = f^{ij}_{\ \ k} T^k, \label{Nb2}\\
&[T^i, T^j, T^k] = f^{ijk} T^{-1}, \label{Nb3}
\end{align}
where $a,b = \{-1, 0, i \}$. $T^{i}$ are generators
of the ordinary Lie algebra with the structure
constant: $[T^i, T^j]=if^{ij}_{\ \ k}T^k$.
This 3-algebra satisfies the fundamental identity.
The metric $h^{ab}=\, \tr \,(T^a, T^b)$  is given by
\begin{align}
 &\, \tr \,(T^{-1}, T^{-1}) = \, \tr \, (T^{-1}, T^{i}) = 0, \ \
  \, \tr \,(T^{-1},T^0) = -1, \nonumber \\
 &\, \tr \,(T^0,T^i) = 0, \ \ \, \tr \,(T^0, T^0) = 0, \ 
 \, \tr \,(T^i,T^j) = h^{ij}.
\end{align}
Since the metric has a negative eigenvalue, the 
field associated with the generators $T^{-1}$
and $T^0$ become ghost modes. 

The gauge generators of the Lorentzian 3-algebra
can be classified into 3 classes:
\begin{itemize}
\item{${\cal I}$}=$\{ T^{-1} \otimes T^a , a=0,i\}$
\item{${\cal A}$}=$\{T^0 \otimes T^i  \}$
\item{${\cal B}$}=$\{ T^i \otimes T^j \}$.
\end{itemize}
The generators in the class ${\cal I}$
vanish when they act on $X$,
hence we  set these generators to zero in the following. 
Since the generators in the class ${\cal B}$ always
appear as a combination with the structure constant,
we define generators $S^i\equiv f^i_{jk} \tilde{T}^{jk}$.
Then they satisfy the algebra
\begin{align}
 [\tilde{T}^{0i}, \tilde{T}^{0j}] = i f^{ij}_k \tilde{T}^{0k},
\ \ 
 [\tilde{T}^{0i}, S^j] = i f^{ij}_k S^k, \ \ 
 [S^i, S^j]=0.
 \label{BLalgebra}
\end{align}
The last commutator was originally proportional 
to the generators in the class ${\cal I}$.
If we had kept these generators, the algebra
would have become  nonassociative.
The algebra (\ref{BLalgebra})
is a semi direct sum of 
$SU(N)$ (or $U(N)$) and translations.
In the case of $SU(2)$, it becomes  the $ISO(3)$
gauge group, which is the gauge group of the 
3-dimensional gravity.
The Lorentzian Bagger-Lambert theories have the
above gauge symmetries and corresponding gauge
fields $\hat{A}_\mu$ and $B_\mu$ as we will see in the next section.

On the other hand, the theory proposed by Aharony
et.al. \cite{Aharony:2008ug} is a Chern-Simons (CS) gauge theory with 
the gauge group $U(N) \times U(N)$.
They act on the bifundamental fields (e.g. $X^I$)
from the left and the right as 
$X \rightarrow UXV^\dagger$.
If we write the generators  as 
$T_L^i$ and $T_R^i$, the combination
$T^i =T_L^i+ T_R^i$ and $S^i=T_L^i- T_R^i$
satisfy the algebra
\begin{equation}
[T^i,T^j]=i f^{ij}_k T^k, \ \ 
[T^i, S^j]=i f^{ij}_k S^k, \ \
[S^i, S^j]= i f^{ij}_k T^k. 
\label{ABJMalgebra}
\end{equation}
By taking the In\"on\"u-Wigner contraction, i.e.
scaling the generators as 
$S^i \rightarrow \lambda^{-1} S^i$ and taking 
$\lambda \rightarrow 0$ limit, 
the algebra (\ref{ABJMalgebra}) becomes
the algebra (\ref{BLalgebra}) of the Lorentzian
BL theory. Therefore it is tempting to think
that the Lorentzian BL theory can be obtained 
by taking an appropriate scaling limit of the 
ABJM theory.
In the next section, we see that it is indeed the
case. Interestingly, even the constraint 
equations in the BL theory (obtained by integrating
the Lagrange multiplier fields) 
can be derived from this scaling procedure.


\section{Derivation of  Lorentzian BL from ABJM}
\setcounter{equation}{0}
\subsection{Lorentzian BL theory}
We first give a quick summary of the Lorentzian BL theory.
Bagger-Lambert theory
 is a (2+1)-dimensional non-Abelian gauge theory with ${\cal N}=8$ 
supersymmetries. It contains 8 real scalar fields 
$X^I=\sum_a X^I_a T^a, \ I=3,...,10$,
gauge fields $A^\mu=\sum_{ab}A^\mu_{ab} T^a 
\otimes T^b, \mu=0,1,2$ with two internal indices
and 11-dimensional Majorana spinor fields 
$\Psi= \sum_a\Psi_a T^a$ with a chirality
condition $\Gamma_{012}\Psi=-\Psi.$
The action proposed by Bagger and Lambert 
is given by
\begin{align}
{\cal L} = -\frac{1}{2} \, \tr \,(D^{\mu}X^I, D_{\mu} X^I) 
           + \frac{i}{2} \, \tr \,(\bar\Psi, \Gamma^{\mu}D_{\mu}\Psi) 
           +\frac{i}{4} \, \tr \,(\bar\Psi, \Gamma_{IJ} [X^I, X^J, \Psi])
           -V(X) + {\cal L}_{CS}. \label{BLaction}
\end{align}
where
$D_{\mu}$ is the covariant derivative defined by
\begin{align}
 (D_{\mu} X^I)_a = \partial_{\mu} X^I_a - f^{cdb}_{\ \ \ a} A_{\mu cd}(x) X^I_b.
\end{align}
$V(X)$ is a the sextic potential term 
\begin{align}
 V(X) = \frac{1}{12} \, \tr \,([X^I,X^J,X^K],[X^I,X^J,X^K]),
\end{align}
and the Chern-Simons term for the gauge potential is given by
\begin{align}
 {\cal L}_{CS} = \frac{1}{2} \epsilon^{\mu\nu\lambda} (f^{abcd} A_{\mu
  ab} \partial_{\nu} A_{\lambda cd} + \frac{2}{3}f^{cda}_{\ \ \ g} f^{efgb}
  A_{\mu ab} A_{\mu cd} A_{\lambda ef}).
\end{align}

In the specific realization of the 3-algebra generated by 
$(T^{-1}, T^0, T^i)$, 
we can decompose the modes of the fields as
\begin{align}
 X^I &= X^I_0 T^0 + X^I_{-1} T^{-1} + X^I_i T^i, \\
\Psi &= \Psi_0 T^0 + \Psi_{-1} T^{-1} + \Psi_i T^i, \\
A_{\mu} &= T^{-1} \otimes A_{\mu(-1)} - A_{\mu(-1)} \otimes T^{-1}
 \nonumber \\ &  + A_{\mu 0j} T^0 \otimes T^j - A_{\mu j0}
 T^j \otimes T^0 + A_{\mu ij} T^i\otimes T^j.  
\end{align} 
It will be convenient to define the following fields as in \cite{Ho:2008ei}
\begin{align}
 &\hat{X^I} = X^I_{i} T^i, \hspace{2em}
  \hat{\Psi} = \Psi_{i} T^i \nonumber \\
 &\hat{A}_{\mu} = 2 A_{\mu 0 i} T^i, \hspace{1em} B_{\mu} =
  f^{ij}{}_k A_{\mu ij} T^k.
\end{align}
The gauge field $A_{\mu (-1)}$ is decoupled from
the action and we drop it in the following discussions.
The gauge field $\hat{A}_{\mu}$ is associated with the 
gauge transformation of the subalgebra ${\cal A}$.
Another gauge field $B_\mu$ will play a role of the 
$B$-field of the BF theory.
With these expression the Bagger-Lambert action (\ref{BLaction}) can be
rewritten as ${\cal L}_{BL}={\cal L}_0+{\cal L}_{gh}$ where
\begin{align}
 {\cal L}_0 &= \, \tr \,
   \left[ - \frac{1}{2}(\hat{D}_{\mu} \hat{X}^I - B_{\mu} X_0^I)^2
    + \frac{1}{4}(X_0^K)^2 ([\hat{X}^I,\hat{X}^J])^2
    - \frac{1}{2} (X_0^I [\hat{X}^I,\hat{X}^J])^2
    \right. \nonumber \\
  & \hspace{2em} 
   + \frac{i}{2} \bar{\hat{\Psi}} \Gamma^{\mu} \hat{D}_{\mu} \hat{\Psi} +
   i \bar{\Psi}_0 \Gamma^{\mu} B_{\mu} \hat{\Psi} 
      -\frac{1}{2}\bar{\Psi}_0 \hat{X}^I
      [\hat{X}^J,\Gamma_{IJ}\hat{\Psi}]
      + \frac{1}{2}\bar{\hat{\Psi}}X^I_0[\hat{X}^J,\Gamma_{IJ}\hat{\Psi}]
          \nonumber \\
  &  \hspace{19em}
   \left.
    + \frac{1}{2} \epsilon^{\mu\nu\lambda} \hat{F}_{\mu\nu} B_{\lambda}
      - \partial_{\mu} X^I_0\  B_{ \mu} \hat{X}^I
 \right],
 \label{L-BLaction}
\end{align}
and
\begin{align}
 {\cal L}_{gh} =   (\partial_{\mu} X^I_0)(\partial^{\mu} X^I_{-1}) - i \bar{\Psi}_{-1}
  \Gamma^{\mu} \partial_{\mu} \Psi_0.
\end{align}
The fields  $X_{-1}$ and $\Psi_{-1}$ are contained in ${\cal L}_{gh}$ only.

The covariant derivative and the field strength
\begin{align}
 \hat{D}_{\mu} \equiv \partial_{\mu} \hat{X}^I +i 
  [\hat{A}_{\mu},\hat{X}^I], \hspace{1em}
  \hat{D}_{\mu} \Psi \equiv \partial_{\mu} \hat{\Psi} +i 
  [\hat{A}_{\mu},\hat{\Psi}], \hspace{1em} \hat{F}_{\mu\nu} =
  \partial_{\mu}\hat{A}_{\nu} - \partial_{\nu} \hat{A}_{\mu} +i 
  [\hat{A}_{\mu}, \hat{A}_{\nu}]
\end{align}
are the ordinary covariant derivative and field strength for the 
subalgebra ${\cal A}.$

Here note that 
$X^I_{-1}$ and $\Psi_{-1}$ appear only linearly in the  Lagrangian
and thus they are  Lagrange multipliers. 
By integrating out these fields, we have the following 
constraints:
\begin{align}
 \partial^2 X^I_0 = 0, \hspace{2em} \Gamma^{\mu} \partial_{\mu} \Psi_0 = 0.
\label{constraints}
\end{align}
Then the Lorentzian BL theory Lagrangian is described by the Lagrangian ${\cal L}_0$.
The constraint equations (\ref{constraints}) and the 
Lagrangian ${\cal L}_0$
are what we want to obtain from the ABJM theory by
taking a scaling limit.
\subsection{Classical Conformal symmetry of D2-branes with dynamical coupling}
Before discussing the ABJM theory, we investigate the symmetry 
properties of the Lorentzian BL theory.
As was shown in \cite{Gomis:2008uv,Benvenuti:2008bt,Ho:2008ei},
the theory can be reduced to a system of D2-branes by integrating
$B_\mu$ fields.
This is interpreted as giving a VEV to $X_0^I$ field
following \cite{Mukhi:2008ux}, and a special
solution $X_0^I=const.$ to  the constraint equation 
$\partial^2 X_0^I =0$  was considered.
In our previous paper \cite{Honma:2008un}, we 
revisited the constraint equation and considered 
a general solution with space-time dependent 
$X_0^I(x)$ satisfying $\partial^2 X_0^I=0.$
Our interpretation is slightly different from the original one, 
and  the field $X_0^I$ is treated as a  dynamical
(but nonpropagating) field.
In this subsection we show that if we consider the whole
set of the solutions to the constraint equation
the reduced action has a classical conformal symmetry 
as well as $SO(8)$ symmetry.

For simplicity, we neglect the fermionic field here.
By integrating the  $B_\mu$ gauge field the action becomes \cite{Honma:2008un}
\begin{align}
{\cal S}_{0} = \int d^3x \ 
\, \tr \, \left[  -\frac{1}{2} (\hat{D}_{\mu} Y^I)^2 + \frac{1}{4} X_0^2
  [Y^I, Y^J]^2 
-\frac{1}{4 (X_0)^2} \big( 
\hat{F}_{\mu \nu} + 2 \epsilon_{\mu \nu \rho}Y_I 
 \partial^\rho  X_0^I  \big) ^2 
 \right],
\label{JanusFT0}
\end{align}
where $X_0^2 \equiv \sum_{I} X_0^I X_0^I$ and
we have defined a new scalar field $Y^I = P_{IJ} \hat{X}^J$
with  7 degrees of freedom by using the projection 
operator
\begin{equation}
P_{IJ}(x) = \delta_{IJ} - \frac{X_{0I} X_{0J}}{X_0^2}.
\end{equation}
Indices run $I,J = 0,\cdots, 8$ and $Y^I$ transforms as a vector of $SO(8)$.
The field $X_0^I(x)$  is constrained to satisfy $\partial^2 X_0^I=0.$
If we pick up a specific solution $X_0^I=v \delta^I_{10}$, 
the action is reduced to 
the familiar D2-brane effective action with a coupling 
constant given by $v$. 
Then $SO(8)$ symmetry is spontaneously broken to $SO(7)$.
The conformal invariance is also broken\footnote{
In the paper \cite{Gomis:2008be}, 
it is discussed that 
the conformal invariance can be  restored by 
sending the Yang-Mills coupling
to infinity or integrating it over all values.}.
However if we consider whole set of solutions,
$SO(8)$ invariance is restored in the action~(\ref{JanusFT0}) with the background fields $X_0^I(x)$ 
although 
$Y^I$ has only 7 degrees of freedom.

Another important symmetry of the action is a 
conformal symmetry. 
The ordinary D2-brane action with a fixed coupling constant
is not conformally invariant and the near horizon limit is
not described by the AdS geometry.
However, as discussed in a paper by Jevicki, Kazama and Yoneya
\cite{Jevicki}, Dp brane theory has a 
{\it generalized conformal symmetry} 
if the coupling   $g(x)$ is not constant and varies with space-time.
Our reduced action for D2-branes (\ref{JanusFT0}) 
has exactly the same property.
The coupling constant is no longer a constant and varies with space-time.
A big difference, however, is that in our case the coupling constant
$g$ is promoted to an $SO(8)$ vector $X_0^I$, which
 is  a space-time dependent field satisfying
the massless Klein-Gordon equation. 

Under the dilation $x \rightarrow \exp(\epsilon) x$,
each field transforms as 
$Y(x) \rightarrow Y'(x')= \exp(-\epsilon/2) Y(x)$, 
 $X_0(x) \rightarrow X_0'(x')=\exp(-\epsilon/2) X_0(x)$
and $A_\mu(x) \rightarrow A_\mu'(x') = \exp(-\epsilon) A_\mu(x)$.
It is easy to see that the action is invariant under the dilation.
Special conformal transformations are more complicated.
It is given by 
\begin{equation}
\delta \ x^\mu = 2 \epsilon \cdot x x^\mu - \epsilon^\mu x^2.
\end{equation}
Writing  an infinitesimal transformation for each field
as $\delta Y(x)=Y'(x')-Y(x)$, we define a
 special conformal transformation for each field as\footnote{
 If $X_0^I$ is replaced by a single field $g(x)$, the transformation
 is the same as the generalized conformal transformation in \cite{Jevicki}.
 Our scalar field $Y(x)$ corresponds to their  $X(x)/g(x)$ .}
\begin{align}
\delta Y^I(x) &= -\epsilon \cdot x Y^I(x) \\
\delta X_0^I(x)&=  -\epsilon \cdot x X_0^I(x)
 \label{conformal transformation of X_0}\\
\delta A_\mu(x) &= -2 \epsilon \cdot x A_\mu(x) -
 2(x\cdot A \ \epsilon_\mu - \epsilon \cdot A \ x_\mu ).
\end{align} 
It is straightforward to show that the action is invariant
under the special conformal transformation\footnote{
The kinetic term of the gauge fields is different
from the ordinary one, but both of the ordinary 
type and ours are invariant under
the same conformal transformations.}. 
It can be easily checked that the transformations preserve the
condition $X_0 \cdot Y=0$.

Finally we need to check that the transformation is closed within
the constraint equation $\partial^2 X_0^I=0$.
From the transformation of $X_0^I$, we  define the 
following transformation
{\it at the numerically same point} as
\begin{equation}
\tilde{\delta}X_0(x) = X_0'(x) -X_0(x) = \delta X_0(x) - \delta x^\mu \partial_\mu X_0(x).
\end{equation}
It is easy to see that if 
the original $X_0^I(x)$ satisfies the constraint equation 
$\partial^2 X_0(x)=0$, then 
the infinitesimal variation satisfies $\partial^2 (\tilde{\delta}X_0)=0$
for both of the dilation and the special conformal transformations,
which means that the transformed field also satisfies
$\partial'^2 X_0'(x')=0.$
Hence the  classical conformal transformation is closed 
within the configurations of $X_0$ satisfying the constraint
equation
\footnote{
In order to construct a set of solutions in which the conformal transformations
are closed, it seems to be necessary to consider all the solutions
to the constraint equation $\partial^2 X_0=0$.
Instead we can consider the following set of solutions studied by Verlinde
\cite{Verlinde:2008di} 
\[
 X_0^I(x) =\sum_{i}\frac{q^I_i}{|x-z_i|} 
\]
which satisfies the constraint equation with sources at $x=z_i$:
 $\partial^2 X_0^I = -4\pi\sum q^I_i \delta^3(x-z_i)$.
 These  solutions are closed under  the conformal transformation
 if we consider all set of $q_i^I$ and $z_i$.
See \cite{Honma:2008ef} for details.
}
.  
If we restrict the configurations of $X_0^I$ that satisfy $\partial X_0^I=0$,
namely, to a set of constant vectors,
the above special conformal transformations cannot be defined
within the set. This indicates that taking into account the whole set of 
the constraint equation  $\partial^2 X_0^I=0$ as adopted in 
\cite{Honma:2008un} is  important  in  
recovering  the $SO(8)$ superconformal symmetry.
It is also interesting to note that $\partial^2 (\tilde{\delta}X_0)=0$
holds only when $p=2$. (Generalized conformal transformations
for general $p$ are given in \cite{Jevicki}.)

As we see later, the D2-brane action with a space-time dependent coupling
is also derived from the M2-brane theory given by Aharony et.al
by taking a certain scaling limit.
This scaling limit corresponds to locating the M2-branes
far from the origin of the orbifold and then taking the $k \rightarrow \infty$
limit. It is  natural from this brane picture that 
the model we considered in this subsection has a 
classical conformal symmetry as well as $SO(8)$ symmetry. 

More detailed studies of the conformal 
symmetries  and the interpretation in 
the gravity side
are discussed in a separate paper \cite{Honma:2008ef}.
What we have suggested here is that if we allow the background fields $X_0^I$ to transform under $SO(8)$ and a special conformal transformation as an $SO(8)$ vector and as in (\ref{conformal transformation of X_0}), the action (\ref{JanusFT0}) is invariant under them.
Note also that the analysis here is just about the classical conformal invariance.
It is interesting to see whether the conformal invariance
can be preserved quantum mechanically.

\subsection{ABJM theory}
The action of the ABJM theory is given by (we use the convention
used in \cite{Benna:2008zy})
\begin{align}
 {\cal S} &= \int d^3 x \ \, \tr \, [- (D_{\mu}Z_A)^{\dagger}D^{\mu} Z^A
  - (D_{\mu} W^A)^{\dagger} D^{\mu} W_A + i \zeta^{\dagger}_A
  \Gamma^{\mu} D_{\mu} \zeta^A + i \omega^{\dagger A} \Gamma^{\mu}
  D_{\mu} \omega_A ] \nonumber \\
 & \quad  + {\cal S}_{CS} - {\cal S}_{V_f} - {\cal S}_{V_b},
\end{align}
with $A = 1,2$. This is an ${\cal N} = 6$ superconformal $U(N) \times
U(N)$ Chern-Simons theory. 
$Z$ is a bifundamental field under the gauge group and its
 covariant derivative is defined by
\begin{align}
 D_{\mu} X = \partial_{\mu} X + iA^{(L)}_{\mu} X - iX A^{(R)}_{\mu}.
\end{align}
The gauge transformations $U(N) \times U(N)$ act from the left
and the right on this field as $Z \rightarrow UZV^\dagger$.

The level of the Chern-Simons gauge theories is $(k,-k)$ and
the coefficients of the Chern-Simons terms  for the
two $U(N)$ gauge groups, 
$A^{(L)}_{\mu}$ and $A^{(R)}_{\mu}$, are opposite.
Hence the action ${\cal S}_{CS}$ is given by
\begin{align}
 {\cal S}_{CS} = \int d^3 x \ 2K \epsilon^{\mu\nu\lambda}
  \, \tr \, [A^{(L)}_{\mu}\partial_{\nu} A^{(L)}_{\lambda} + \frac{2i}{3}
  A^{(L)}_{\mu}A^{(L)}_{\nu}A^{(L)}_{\lambda} 
  - A^{(R)}_{\mu}\partial_{\nu}
  A^{(R)}_{\lambda} - \frac{2i}{3}
  A^{(R)}_{\mu} A^{(R)}_{\nu} A^{(R)}_{\lambda}].
\end{align}
The potential term for bosons is given by
\begin{align}
 S_{V_b} &= -\frac{1}{48 K^2} \int d^3 x \ \, \tr \,[
  Y^A Y^{\dagger}_A Y^B Y_B^{\dagger} Y^C Y^{\dagger}_C
  + Y^{\dagger}_A Y^A Y^{\dagger}_B Y^B Y_C^{\dagger} Y^C \nonumber \\
  &  \ \ \ \ \ \ \ \ \ \ \ \ \ \ \ \ \ \ \ \ 
   + 4 Y^A Y^{\dagger}_B Y^C Y^{\dagger}_A Y^B Y_C^{\dagger}
   - 6 Y^A Y^{\dagger}_B Y^B Y^{\dagger}_A Y^C Y_C^{\dagger}],
\end{align}
and for fermions by
\begin{align}
 S_{V_f} &= \frac{i}{4K} \int d^3 x \, \tr \, [Y^{\dagger}_A Y^A
  \psi^{B\dagger} \psi_B - Y^A Y^{\dagger}_A \psi_B \psi^{B\dagger} + 2
  Y^A Y^{\dagger}_B \psi_A \psi^{B\dagger} - 2 Y^{\dagger}_A Y^B
  \psi^{A\dagger} \psi_B \nonumber \\
 &  \ \ \ \ \ \ \ \ \ \ \ \ \ \ \ \ 
  + \epsilon^{ABCD} Y^{\dagger}_A \psi_B Y^{\dagger}_C \psi_D -
  \epsilon_{ABCD} Y^A \psi^{B\dagger} Y^C \psi^{D\dagger}].
\end{align}
$Y^A$ and $\psi_A$ ($A=1 \cdots 4)$ are defined by
\begin{align}
 Y^C = \{ Z^A, W^{\dagger A} \}, \ \ \ \ \ \ \psi_C = \{\epsilon_{AB}
  \zeta^B e^{i\pi/4}, \epsilon_{AB} \omega^{\dagger B} e^{-i\pi/4}\},
\end{align}
where the index $C$ runs from $1$ to $4$. The $SU(4)$ R-symmetry 
of the potential terms is
manifest in terms of $Y^A$ and $\psi_A$. 

The ABJM theory is similar to the Lorentzian BL theory, but
different in the following points. First the gauge group is 
$U(N) \times U(N)$ while it is a semi direct product of $U(N)$
and translations in the BL theory.
Accordingly the matter fields
are in the bifundamental representation in the ABJM theory.
Furthermore the BL theory contains an extra field $X_0$ 
and $\Psi_0$ associated with the generator $T_0$, and 
they are required  to obey the constraint equations (\ref{constraints}).

The bosonic potential terms in both theories are sextic, but 
the potential in the BL theories contains two $X_0^I$ fields and 
four adjoint matter fields $\hat{X}^I$ while the potential terms 
in the ABJM theory are written in the product of six bifundamental
matter fields $Y$. Hence it is natural to think that 
the trace part of $Y$ will play a role of $X_0$ in the 
Lorentzian BL theory.
We will see that, 
if we separate the matter field $Y$ into
a trace and a traceless part, the  potential terms  
coincide in a certain scaling limit.

\subsection{Scaling limit of ABJM theory}
In order to take a scaling limit,  we first recombine the gauge fields as
\begin{align}
 \hat{A}_{\mu} = \frac{A^{(L)}_{\mu} + A^{(R)}_{\mu}}{2}, \ \ \ \ \
  B_{\mu} = \frac{A^{(L)}_{\mu} - A^{(R)}_{\mu}}{2},
\end{align}
then the gauge transformations corresponding to $\hat{A}_{\mu}$ and $B_{\mu}$
are $Z \rightarrow e^{i\sigma_a T^a}Z e^{-i\sigma_b T^b}$ and
$Z\rightarrow e^{i \sigma_a T^a} Z e^{i \sigma_b T^b}$ respectively. 
They are vectorial and axial gauge transformations.
Matter fields are in the adjoint representation for the $\hat{A}_\mu$ gauge fields.
Hence the 
$U(1)$ part of  $\hat{A}_{\mu}$ decouples from the matter sector.

The covariant
derivative can be written in terms of $\hat{A}_{\mu}$ and $B_{\mu}$ as
\begin{align}
 D_{\mu} Z &= \partial_{\mu} Z + i[\hat{A}_{\mu},Z] + i\{ B_{\mu}, Z \}
  \nonumber \\
 &= \hat{D}_{\mu} Z + i\{B_\mu, Z\},
 \label{CDer}
\end{align}
where $\hat{D}_{\mu}$ is the covariant derivative with respect to
the gauge field $\hat{A}_{\mu}$. 
$S_{CS}$   can be written in terms of $\hat{A}_{\mu}$ and $B_{\mu}$ as
\begin{align}
 S_{CS} = \int d^3 x \ 4K \epsilon^{\mu\nu\rho} \, \tr \,
  [B_{\mu} \hat{F}_{\mu\nu} + \frac{2}{3} B_{\mu}B_{\nu}B_{\rho}],
\end{align}
where $\hat{F}_{\mu\nu}$ is field strength of $\hat{A}_{\mu}$. \\
The gauge fields $\hat{A}_{\mu}$, $B_{\mu}$ are associated with 
the gauge transformations generated by $T^i$ and $S^i$
in (\ref{ABJMalgebra}). Hence in order to take the In\"on\"u-Wigner
contraction to obtain the gauge structure of the Lorentzian
BL theory (\ref{BLalgebra}), we need to rescale the gauge field
$B_\mu$ as  $B^\mu \rightarrow \lambda B^\mu$ and take the $\lambda \rightarrow 0$ limit.
Simultaneously we need to scale the coefficient $K$ by $\lambda^{-1} K$.
Since the coefficient $K$ is proportional to the level of the Chern-Simons theory
$k$ as $K=k/8\pi$, the scaling limit corresponds to taking
the large $k$ limit.
In this scaling limit, the cubic term of the 
$B_\mu$ fields
vanishes and the 
Chern-Simons action coincides with the BF-type action in the 
Lorentzian BL theory:
\begin{equation}
 S_{CS} \rightarrow  \int d^3 x \ 4K \epsilon^{\mu\nu\rho} \, \tr \,
  B_{\mu} \hat{F}_{\mu\nu}. 
  \end{equation}

In order to match 
 the covariant derivatives in the Lorentzian BL action (\ref{L-BLaction})
and in the ABJM theory (\ref{CDer}), 
we  separate the bifundamental fields into the
trace  and the traceless part, and scale them differently.
We write the matter fields $Y^A$  as
\begin{align}
 Y^A_{ij} = Y^A_0 \delta_{ij} + \tilde{Y}^{A}_a T^a_{ij},
\end{align}
where $T^a$ is the generator of $SU(N)$.

Now we perform the following rescaling:
\begin{align}
 B_{\mu} &\rightarrow \lambda B_{\mu}, \nonumber \\
 Y^{A}_0 &\rightarrow \lambda^{-1} Y^A_0, \nonumber \\
 \psi_{A0} &\rightarrow \lambda^{-1} \psi_{A0}, \nonumber \\
 K &\rightarrow \lambda^{-1}K,
\end{align}
where $Y_0^A$ and $\psi_{A0}$ is the trace part of $Y^A$ and
$\psi_A$.  All the other fields are kept fixed. 
Then take the $\lambda \rightarrow 0$ limit. 
If we take the scaling limit, we can show that the covariant derivatives
in both theories exactly match.

In the following we consider the ABJM theory with the $SU(N) \times SU(N)$
gauge group. In the presence of the $U(1)\times U(1)$ group,
a little more care should be taken for the scaling of the
$U(1)$ part of the $B_\mu$ gauge field.

In taking the above scaling limit, 
many terms vanish.  The kinetic term of the 
ABJM action becomes
\begin{align}
 &  \, \tr \, \left[-\frac{1}{\lambda^2} \partial_{\mu}
  Y_{0A} ^{\dagger} \partial^{\mu} Y^A_0 
  + \frac{1}{\lambda^2} \psi^{\dagger}_{0A} \Gamma^{\mu}
  \partial_{\mu} 
  \psi^A_0 + 2 (i\partial_\mu Y_{0A}^\dagger B^\mu Y^A +h.c.) \right.
\nonumber \\
 &  \left.\hspace{0.5em}
  - (\hat{D}_{\mu} \tilde{Y}_{A} + 2i \tilde{B}_{\mu} Y_{0A})^{\dagger}
  (\hat{D}^{\mu} \tilde{Y}^A + 2i \tilde{B}^{\mu} Y^A_0)
  + i \tilde{\psi}_A^{\dagger} \Gamma^{\mu} \hat{D}_{\mu} \tilde{\psi}^A 
  -2\tilde{\psi}^{\dagger}_A \Gamma^{\mu} \tilde{B}_{\mu} \psi^A_0 -
  2\psi^{\dagger}_{0A} \Gamma^{\mu} \tilde{B}_{\mu} \tilde{\psi}^A
  \right]. \notag\\
 \label{kinetic}
\end{align}
The first and the second terms are divergent for small $\lambda$.
In order to make the action finite, 
we need to impose that
the trace part of the bifundamental fields must 
satisfy the constraint equations
\begin{equation}
\partial^{2} Y_0^I = 0, \ \ 
\Gamma^{\mu} \partial_{\mu} \psi_{A0} = 0\notag
\end{equation}
in the $\lambda \rightarrow 0$ limit.
They are precisely the same constraint equations 
(\ref{constraints}) in the BL theory.\\
In the Lorentzian BL theory, the constraints are obtained
by integrating out the Lagrange multiplier
 fields $X_{-1}$ and $\Psi_{-1}$.
Here they arise from a condition that the action should be finite
in the scaling limit. 

The other terms in (\ref{kinetic}) are finite in the 
scaling limit and it can be easily shown that they are precisely 
the same
kinetic terms as that of  the Lorentzian Bagger-Lambert theory
(after a redefinition of the gauge field 
$2B_{\mu} \rightarrow B_{\mu}$ and setting $K=1/2$).
The trace part of the bifundamental fields is identified
with the fields  $X_0$ associated with one of the extra generators $T^0$
in the Lorentzian Bagger-Lambert theory.
This is the reason why we have used the same convention with
subscript $0$ for both of 
the trace part of the bifundamental fields
and the field associated with the generator $T^0.$

Now let us check the potential terms. 
The potential terms of the ABJM theory are invariant under
the $SU(4)$ symmetries but not under full $SO(8)$. 
By decomposing the matter fields $Y^A$
into the trace part $Y^A_0$ and the traceless part
$\tilde{Y}^A$, the bosonic sextic potential becomes
a sum of $V_B=\sum_{n=0}^{6} V_B^{(n)}$, where $V_B^{(n)}$
contains $n$ $Y_0$ fields and $(6-n)$ $\tilde{Y}$ fields.
Since the coefficient of the bosonic potential is proportional 
to $K^{-2}$, $V_B^{(n)}$ term scales as $\lambda^{2-n}$.
It can be easily checked that the coefficients of 
$V_B^{(n)}$ vanishes for $n >3$.
On the other hand, the potential terms $V_B^{(n)}$ for $n<2$ vanish
in the scaling limit of $\lambda \rightarrow 0$.
Hence the only remaining term in the scaling limit is $V_B^{(2)}$.
This part of the potential has the full $SO(8)$
symmetry and becomes identical with the potential in the Lorentzian
BL theory. 
In order to see that  the BL potential is obtained, 
we  assume that only the field $Z^1$ has 
the trace part for simplicity.
Let us write the $4$ complex scalar field $Y^A$
by 8 real scalar fields as
\begin{align}
 Z^1 &= X^1_0 + iX^5_0 + i\tilde{X}^1_a T^a - \tilde{X}^5_a T^a,
  \nonumber \\
 Z^2 &= i\tilde{X}^2_a T^a - \tilde{X}^6_a T^a,  \nonumber \\
 W_1^{\dagger} &= i\tilde{X}^3_a T^a - \tilde{X}^7_a T^a \nonumber \\
 W_2^{\dagger} &= i\tilde{X}^4_a T^a - \tilde{X}^8_a T^a.
 \label{boson components}
\end{align}
Substituting them into $S_{V_b}$ and taking the scaling 
limit, we can  obtain the following bosonic potential:
\begin{align}
 S_{V_b} = - \frac{1}{8K^2} \int d^3 x  \ \, \tr \,
\left( (X^1_0)^2+(X^5_0)^2)
  [P_I,P_J][P^I,P^J] \right).
\end{align}
  $P^I$ is defined by
\begin{align}
 P^I &\equiv (P^1,\tilde{X}^2,\tilde{X}^3,\tilde{X}^4,\tilde{X}^6,
  \tilde{X}^7,\tilde{X}^8),\notag\\
 &= \left(\frac{1}{2}(\tilde{Y}^A + \tilde{Y}_A^\dagger),
     \frac{1}{2i}(\tilde{Y}^B - \tilde{Y}_B^\dagger)\right),\\
 \tilde{Y}^A &\equiv (P^1, Z^2,W_1^\dagger,W_2^\dagger),\notag\\
 P^1&\equiv\frac{X^1_0\tilde{X}^5-X^5_0\tilde{X}^1}{\sqrt{(X^1_0)^2+(X^5_0)^2}}  .\notag
 \label{definition of P^I}
  \end{align}
We can rewrite it as,
\begin{align}
 S_{V_b} = - \frac{1}{8K^2}\int d^3 x \,\tr \,\left[
  \frac{1}{4}(X_0^K)^2 \left([\tilde{X}^I,\tilde{X}^J]\right)^2
  - \frac{1}{2}\left(X_0^I [\tilde{X}^I,\tilde{X}^J]\right)^2\right],
\end{align}
where we have used ${X}_0^I = (X_0^1,0,0,0,X_0^5,0,0,0)$.
This is the potentials for bosons in the Lorentzian BL theory~(\ref{L-BLaction}).
It is straightforward to see that the complete potential 
of the BL theory can be obtained  by considering 
general $X_0^I$  and  
the full $SO(8)$ invariance is restored.

 It should be noted that the above potential term is written in terms of the
 commutators. This shows that, if we replace more than two 
  bosons by their trace components, the potential  vanishes.
  This assures that the would-be divergent terms 
  $V_B^{(n)}$  for $n>3$  vanish
and the only remaining  term in the scaling limit
is given by the above   potential.

Finally  consider the fermion potential.
We expand the potential as $V_f=\sum_{n=0}^{4} V_f^{(n)}$
where $V_f^{(n)}$ contains $n$ trace parts and $(4-n)$ traceless parts.
Since the coefficient of the fermion potential is proportional to
$1/K$,  $V_f^{(n)}$ scales as $\lambda^{1-n}$. 
$V_f^{(n)}$ for $n>1$ diverges in the scaling limit and their
coefficients must vanish.
$V_f^{(0)}$ vanishes in the scaling limit $\lambda \rightarrow 0$.
Hence the only remaining finite terms are $V_f^{(1)}$.
In the following we look at the potential term with one of the 
bosons replaced by the trace part $X_0^I$.
Such a term can be written as 
\begin{align}
 S_{V_f}
  = \frac{i }{2 K} &X_0^1 \,\tr\,\left[
  -\psi_1^\dagger[\tilde{X}^5,\psi_1]+\psi_2^\dagger[\tilde{X}^5,\psi_2]
 +\psi_3^\dagger[\tilde{X}^5,\psi_3] + \psi_4^\dagger[\tilde{X}^5,\psi_4]
 \right. \notag\\
 & \quad
 + \psi_1^\dagger[Y_2,\psi_2]+\psi_2^\dagger[Y_2^\dagger,\psi_1]
 + \psi_3^\dagger[Y_2,\psi_4^\dagger]+\psi_4[{Y_2}^\dagger,\psi_3]
 \notag\\
 & \quad
 +\psi_1^\dagger[Y_3,\psi_3]+\psi_3^\dagger[Y_3^\dagger,\psi_1]
 +\psi_4^\dagger[Y_3,\psi_2^\dagger]+\psi_2[Y_3^\dagger,\psi_4]
 \notag\\
 & \quad \left.
 +\psi_1^\dagger[Y_4,\psi_4]+\psi_4^\dagger[Y_4^\dagger,\psi_1]
 +\psi_2^\dagger[Y_4,\psi_3^\dagger]+\psi_3[Y_4^\dagger,\psi_2]
 \right]\notag\\
 +\frac{i }{2 K} &X_0^5 \,\tr\,\left[
 +\psi_1^\dagger[\tilde{X}^1,\psi_1]-\psi_2^\dagger[\tilde{X}^1,\psi_2]
 -\psi_3^\dagger[\tilde{X}^1,\psi_3] - \psi_4^\dagger[\tilde{X}^1,\psi_4]
 \right. \notag\\
 & \quad
 - \psi_1^\dagger[i Y_2,\psi_2]+\psi_2^\dagger[i Y_2^\dagger,\psi_1]
 + \psi_3^\dagger[i Y_2,\psi_4^\dagger]-\psi_4[i Y_2^\dagger,\psi_3]
 \notag\\
 & \quad
 -\psi_1^\dagger[i Y_3,\psi_3]+\psi_3^\dagger[i Y_3^\dagger,\psi_1]
 +\psi_4^\dagger[i Y_3,\psi_2^\dagger]-\psi_2[i Y_3^\dagger,\psi_4]
 \notag\\
 & \quad \left.
 -\psi_1^\dagger[i Y_4,\psi_4]+\psi_4^\dagger[i Y_4^\dagger,\psi_1]
 +\psi_2^\dagger[i Y_4,\psi_3^\dagger]-\psi_3[i Y_4^\dagger,\psi_2]
 \right].
 \label{ABJM fermion potential}
\end{align}
Here for simplicity we have assumed that the trace part of the boson 
 $X_0^I$ is nonvanishing for $I=1,5$ .
This can be done by using the original $SU(4)$ symmetry.
Note again that these potential terms 
are written as a form of  commutators.

To get the 3-dimensional Majorana fermion as the BL theory, we rewrite the $SU(4)$ 
complex fermion in terms of  the real variables
\footnote{
When we give a VEV to the $X_0^4$ part only, we will get 7 $\Gamma$
matrices as in \cite{Pang:2008hw}.
In our case we need  8 $\Gamma$ matrices  and 
their antisymmetrized-products because
we give a VEV to  a more general direction.}.
\begin{align}
 &\psi_1 = i\chi_1 -\chi_5,\quad \psi_2 = i\chi_2 - \chi_6,\notag\\
 &\psi_3 = i\chi_3-\chi_7,\quad \psi_4 = i \chi_4 - \chi_8,
\end{align}
where $\chi_I$ are  real 2-component spinors.
We also expand the complex bosons as the real ones~(\ref{boson components}).
Then the fermion potential (\ref{ABJM fermion potential})
becomes  by using the
$8\times 8\  \Gamma$ matrice as 
\begin{align}
 S_{V_f}&= -\frac{1}{2K} \,\tr\,
 \Bar{\Psi} X_0^I [\tilde{X}^J ,\Gamma_{IJ} \Psi],\notag\\
 \Psi &\equiv
 \left(\chi_1,\chi_2,\chi_3,\chi_4,\chi_5,\chi_6,\chi_7,\chi_8\right),
 \label{BL fermion potential term}
\end{align}
where the indices $I,J$ run from 1 to 8 and  $X_0^I = (X_0^1,0,0,0,X_0^5,0,0,0)$.
The explicit forms of the $\Gamma$ matrices are given  in the Appendix \ref{sec:gamma-matrices}.
This fermion potential has the same $SO(8)$ invariant form as that of the 
 Lorentzian BL action~(\ref{L-BLaction}).
In the same fashion as the bosonic potential, the full $SO(8)$ invariance can be seen easily by considering the general $X_0^I$.


\section{Conclusions and Discussions}
\setcounter{equation}{0}
In this paper, we have shown that the Lorentzian Bagger-Lambert theory
is derived by taking a scaling limit of the 
${\cal N}=6$ superconformal Chern-Simons field theories proposed by Aharony et.al. 
In the scaling limit the trace components of the matter fields 
are taken to be large compared to the fluctuating traceless components.
Hence the M2-branes are located far from the origin of the 
${\bf C^4/Z_k}$ orbifold (or in the sufficiently low energy). 
Large values of the trace components means that  they have a classical 
VEV and branes  are fluctuating around them.
The $U(N) \times U(N)$ gauge group is broken to the diagonal $U(N)$
and the axial $U(N)$ symmetries become translational symmetries
by the In\"on\"u-Wigner contraction.
Simultaneously in order to keep the action finite we need to
scale the level $k$ of the Chern-Simons gauge theories  to infinity
before taking the large $N$ limit. 
This makes the ${\bf Z_k}$ identification of the orbifold
to become a continuous circle identification.
Since the radius of the M theory is proportional to $1/k$, 
the gravity dual is reduced to $d=10.$
Hence we conclude that the Bagger-Lambert theory based on the 
Lorentzian 3-algebra is a theory of multiple D2-branes. 

However there is a subtlety to interpret the scaled theory as a
$d=10$  IIA superstring on $AdS_4 \times CP^3$.
The Lorentzian BL theory is obtained by the
 $k \rightarrow \infty$ limit before
taking the large $N$ limit. 
The ABJM theory is conjectured to be dual to the M-theory
on $AdS_4 \times S^7/{\bf Z}_k$
and the radius of $AdS _4$ is proportional to 
$R/l_p \sim (kN)^{1/6}$
 in the unit of the Planck scale $l_p$.  
 Hence in $k \rightarrow \infty$  limit 
 the radius becomes large and the $d=11$ supergravity 
 is a good approximation.  
On the other hand,
the compactification radius is 
 proportional to $R/kl_p \propto (Nk)^{1/6}/k$
 and becomes $0$ in the limit. 
Then it is tempting to think that the scaled theory is 
described by type IIA superstring on $AdS_4 \times CP^3$.
However, in the string scale $l_s$
the radius of $AdS_4$ is given by 
$R/l_s \propto (N/k)^{1/4}$ and becomes $0$.
 Hence the gravity dual of the 
scaled theory is more appropriately
interpreted as 
 $d=11$ supergravity in 
$AdS_4 \times CP^3$  space-time rather than a 
type IIA supergravity.

We have discussed that the scaled theory has
 $SO(8)$ invariance if we consider rotations of the background fields $X_0^I(x)$.
Since our scaling limit corresponds to locating the M2-branes far from 
the orbifold singularity, 
the recovery of the $SO(8)$ invariance is natural.
In the field theory side 
we have explicitly checked  the recovery of the  $SO(8)$ invariance 
in the potentials of bosons and fermions. 
It is interesting that the symmetry is enhanced from ${\cal N} = 6$ to ${\cal N}=8$ by 
taking the scaling limit. 

We have also investigated a generalized conformal symmetry of the scaled theory. 
The familiar D2-brane action has a fixed
and space-time independent coupling constant, and
both of the conformal symmetry 
and $SO(8)$ invariance of the M2-branes are broken by the 
VEV of the M2-branes.
On the contrary, our D2-brane action has a 
classical conformal symmetry and $SO(8)$ invariance if we allow the background fields $X_0^I$ to transform under them.
This became possible by promoting the coupling constant
 to a space-time dependent $SO(8)$ vector field  $X_0^I.$ 
Discussions on the gravity side are given in \cite{Honma:2008ef}.
In particular, it is important to clarify where 
we can get the same type of constraint equations  
in the gravity side and see how 
a dual geometry of D2-branes
can  acquire the generalized conformal symmetry. 

\section*{Acknowledgements}
We thank Dr Y. Hikida for useful discussions of 
the ABJM theories and Dr. H. Umetsu for the gauge structures
of the Lorentzian BL theory.

\appendix
\section{The Gamma matrices\label{sec:gamma-matrices}}
The explicit forms of the  antisymmetrized products of
the $8\times 8\ \Gamma$
matrices we have used in~(\ref{BL fermion potential term}) are given as 
$\Gamma_{IJ}=\mathbb{I}_{2\times 2} \otimes \gamma_{IJ}$ where
\begin{align}
 &\gamma_{12} = \left(
 \begin{array}{@{\,}cccc@{\,}}
  i\sigma^2 &&&\\&-i\sigma^2&&\\
  &&i\sigma^2&\\&&&i\sigma^2
 \end{array}
 \right), \quad
 &\gamma_{13} &= \left(
 \begin{array}{@{\,}cccc@{\,}}
  &\mathbb{I}&&\\ -\mathbb{I}&&&\\
  &&&\sigma^3\\&&-\sigma^3&
 \end{array}\right),\notag\\ 
 &\gamma_{14} = \left(
 \begin{array}{@{\,}cccc@{\,}}
  &i\sigma^2&&\\i\sigma^2&&&\\
  &&&\sigma^1\\&&-\sigma^1&
 \end{array}
 \right),\quad
 &\gamma_{15} &= \left(
 \begin{array}{@{\,}cccc@{\,}}
  &&-\sigma^3&\\&&&\mathbb{I}\\
  \sigma^3&&&\\&-\mathbb{I}&&
 \end{array}\right),\notag\\ 
 &\gamma_{16}=\left(
 \begin{array}{@{\,}cccc@{\,}}
  &&-\sigma^1&\\&&&-i\sigma^2\\
  \sigma^1&&&\\&-i\sigma^2
 \end{array}
 \right),\quad
 &\gamma_{17} &= \left(
 \begin{array}{@{\,}cccc@{\,}}
  &&&-\sigma^3\\&&-\mathbb{I}&\\
  &\mathbb{I}&&\\\sigma^3&&&
 \end{array}\right),\notag\\ 
 &\gamma_{18} = \left(
 \begin{array}{@{\,}cccc@{\,}}
  &&&-\sigma^1\\&&i\sigma^2\\
  &i\sigma^2&&\\\sigma^1&&&
 \end{array}\right)\quad,
 &\gamma_{52} &= \left(
 \begin{array}{@{\,}cccc@{\,}}
  &&\sigma^1&\\&&&-i\sigma^2\\
  -\sigma^1&&&\\&-i\sigma^2&&
 \end{array}\right),\notag\\ 
 &\gamma_{53}=\left(
 \begin{array}{@{\,}cccc@{\,}}
  &&&\mathbb{I}\\&&\sigma^3\\
  &-\sigma^3&&\\ -\mathbb{I}&&&
 \end{array}\right),\quad
 &\gamma_{54}&=\left(
 \begin{array}{@{\,}cccc@{\,}}
  &&&i\sigma^2\\&&\sigma^1&\\
  &-\sigma^1&&\\i\sigma^2&&&
 \end{array}\right),\notag\\ 
 &\gamma_{56}=\left(
 \begin{array}{@{\,}cccc@{\,}}
  i\sigma^2&&&\\&i\sigma^2&&\\
  &&i\sigma^2&\\&&&-i\sigma^2\\
 \end{array}\right),\quad
 &\gamma_{57}&=\left(
 \begin{array}{@{\,}cccc@{\,}}
  &\sigma^3&&\\-\sigma^3&&&\\
  &&&\mathbb{I}\\&&-\mathbb{I}&
 \end{array}\right),\notag\\ 
 &\gamma_{58}=\left(
 \begin{array}{@{\,}cccc@{\,}}
  &\sigma^1&&\\-\sigma^1&&&\\
  &&&i\sigma^2\\&&i\sigma^2&
 \end{array}\right) 
 \label{gamma matrices}
\end{align}
and $\mathbb{I}_{2\times 2}$ is a $2 \times 2$ identity  matrix. 
We have also  defined
\begin{align}
 &\Gamma^0 =  i \sigma^2 \otimes  \mathbb{I}_{8\times 8}.
\end{align}
The $i\sigma^2$ was used to contract the indices of
 the 2-component spinor $\chi$
 and it is the 3 dimensional $\gamma^0$ matrix (see the Appendix of \cite{Benna:2008zy}).
$\mathbb{I}_{8\times 8}$ is an $8\times 8$ identity matrix.
They  satisfy the following consistency relations as
$\Gamma_{12} \Gamma_{13}+\Gamma_{13} \Gamma_{12} = -(\Gamma_2\Gamma_3 +
\Gamma_3\Gamma_2) = 0$.
At this stage, there is an ambiguity to determine the
$\Gamma$ matrices, but the explicit forms of $\Gamma_I$ are
not necessary here.
 To fix the ambiguity, we need to
consider more general VEVs of $X_0^I$.

\end{document}